\documentclass[MSNbibl,nameyear,dvips]{arxstspdf}
\usepackage{flushend}
\usepackage{stfloats}

\volume{27}
\issue{2}
\pubyear{2012}
\firstpage{187}
\lastpage{188}
\doi{10.1214/12-STS376A} 
\referstodoi{10.1214/11-STS376}

\begin{document}
\begin{frontmatter}
\vspace*{6pt}
\title{Discussion of ``Statistical Modeling of Spatial Extremes'' by
A.~C. Davison, S.~A.~Padoan and M. Ribatet}
\runtitle{Discussion}

\begin{aug}
\author[a]{\fnms{D.} \snm{Cooley}\corref{}\ead[label=e1]{cooleyd@stat.colostate.edu}}
\and
\author[b]{\fnms{S. R.} \snm{Sain}}
\runauthor{D. Cooley and S. R. Sain}

\address[a]{D. Cooley is Assistant Professor, Department of Statistics, Colorado State University, Fort Collins, Colorado 80523-1877, USA
\printead{e1}.}
\address[b]{Stephan R. Sain is Scientist,
Institute for Mathematics Applied to Geosciences,
National Center for Atmospheric Research,
Boulder, Colorado 80307-3000, USA.}

\end{aug}


\vspace*{-2pt}
\end{frontmatter}

We congratulate the authors for their overview paper discussing
modeling techniques for spatial extremes.
There is great interest in spatial extreme data in the atmospheric
science community, as the data is inherently spatial and it is
recognized that extreme weather events often have the largest economic
and human impacts.
In order to adequately assess the risk of potential future extreme
events, there is a need to know how the characteristics of phenomena
such as precipitation or temperature\break could be altered due to climate change.

Because of the high interest level in the atmospheric science and (more
broadly) the geoscience communities, it is imperative for the
statistics community to develop methodologies which appropriately
answer the questions associated with spatial extreme data.
\citet{davison2011} provide a comprehensive overview of existing
techniques that can serve as a useful starting point for statisticians
entering the field. 
That the paper is written as a case study helps to illustrate the
advantages and disadvantages of the various methods.
We hope that this Swiss rainfall data will serve as a~test set by which
future methodologies can be evaluated.

The authors analyze data which are annual maxima.
This is natural from the classical extreme value theory point of view
whose fundamental result establishes the limiting distribution of $\mathbf
Y = ( \bigvee_{i = 1}^n X_{1i},\break \ldots, \bigvee_{i = 1}^n X_{Di}
)^T$ to be\vadjust{\goodbreak} in the family of the multivariate max-stable distributions.
In practice, modeling vectors of annual maxima seems less than ideal,
and it is not clear how much dependence information is lost by
discarding the coincident data.
Scientists in other disciplines can be uncomfortable with the idea of
constructing data vectors of events which most often occur on different days.
We are aware that there is current work to extend spatial extremes work
to deal with threshold exceedances, and we look forward to that work
appearing in the literature.\looseness=1

\citet{davison2011} divide the spatial approaches into three categories:
latent variable models, copulas, and max-stable process models.
In Section 7 they do a very nice job of detailing the strengths and
weaknesses of the three approaches.
However, it seems that the article does not make clear enough that the
aim of the latent variable approach is fundamentally different than the
aim of a copula or max-stable process model.
As the authors state in Sections 2.2 and 2.3, current modeling of
multivariate (or spatial) extremes requires two tasks: (1) the
marginals must be estimated and transformed to something standard
(e.g., unit Fr\'echet) so that (2) the tail dependence in the data can
be modeled.
The latent variable model is a method for characterizing how the
marginal distribution varies over space, that is, task 1.
In contrast, both copula models and existing max-stable process models
explicitly model the tail dependence in the data once the marginals are
known, that is, task 2.
We refer to the dependence remaining after the marginals have been
accounted for as ``residual dependence,'' as \citet{Sang10} described the
random variables after marginal transformation as ``standardized residuals.''

\citet{davison2011} are correct to point out (Figure 4) that using a
latent variable model is inappropriate for applications where the joint
behavior of the random vector is required.
However, there are applications which aim only to model the marginal behavior.
There is a\vadjust{\goodbreak} long history of producing return level maps such as those
shown in Figure~3 of the manuscript.
For instance, the recent effort to update the precipitation frequency
atlases for the US (\citeauthor{bonnin04a}, \citeyear{bonnin04a,bonnin04b}) aimed only to
characterize the marginal distribution's tail over the study region.
\citeauthor{bonnin04a} (\citeyear{bonnin04a,bonnin04b}) employed a regional frequency analysis
(\cite{dalrymple60}; \cite{hosking97}),  approach which, like the latent variable
model approach, aims to borrow strength across sites when estimating
marginal parameters.
As \citet{davison2011} clearly show, explicitly modeling residual
depend\-ence requires considerable effort, and when only the mar\-ginal
effects need to be described, we feel it can be appropriate to ignore
the residual dependence so long as one recognizes the limited scope of
the questions that such an analysis can answer.

In situations where the joint behavior of multiple locations must be
described, then one must explicitly model the residual dependence.
As \citet{davison2011} show, dependence models not specifically designed
for extremes may be \mbox{inadequate} to capture tail dependence.
However, models such as the extremal copulas or max-stable processes do
not easily lend themselves to current atmospheric science applications
with hundreds, thousands, or tens-of-thousands of locations.
There are obvious avenues to explore toward adapting the pairwise
likelihood methods for large spatial data sets, but, to date, pairwise
likelihood methods have only been applied to applications similar to
the one in \citet{davison2011} with roughly 50 locations.
We imagine scaling the methods to the size of current applications will
be nontrivial, and perhaps new inference procedures or more
computa\-tionally-feasible extremes dependence models will need to be developed.
Until appropriate extremes techniques are available, people will
continue to be tempted to apply high-dimensional models developed to
describe nonextreme data (e.g., a~Gaussian copula) to model tail dependence.

Most of the spatial extremes work to date has been primarily
descriptive in nature.
Such analyses are useful in assessing risk (i.e., the probability of an
extreme event), but do not help to explain the underlying causes of
extreme events.
There is a~desire in the atmospheric sciences to move beyond
descriptive analyses and toward analyses which enhance understanding of
the processes which lead to extreme events.
For example, \citet{sillmann2011} establish a link between extreme cold
temperatures in Europe and\vadjust{\goodbreak} a blocking phenomenon in the North Atlantic,
\citet{maraun2011} link extreme precipitation in Europe to large-scale
airflow covariates, and \citet{weller2011} link extreme precipitation on
the Pacific coast of North America to surface pressure patterns.
Since it is generally believed that climate models are better at
representing processes at large-scales, establishing links between
extreme events and large-scale phenomena enable one to better
conjecture how the nature of extreme events will change with the climate.
While none of the analyses cited above involved extensive spatial
modeling of extremes, it is foreseeable that science will move in this
direction.

Finally, undertaking a pairwise likelihood fitting of a max-stable
process model is challenging and would be beyond the capabilities of
most geoscientists.
The authors are to be commended for developing the \mbox{{\tt
SpatialExtremes}} (\cite{SpatialExtremes}) package in {\tt R} which
enables the general scientific community to utilize these methods.


%

\end{document}